\documentclass[preprint,12pt]{elsarticle}

\usepackage{xspace}
\usepackage{url}
\usepackage{balance}
\usepackage{paralist}

\newcommand{\impact}[4]{\textsf{\small [#1$\,\vert\,$\uppercase{{\scriptsize#2}}] 
$\stackrel{#4}{\longrightarrow}$ [#3]}\xspace}

\newcommand{\impactp}[3]{\impact{#1}{#2}{#3}{+}}
\newcommand{\impactn}[3]{\impact{#1}{#2}{#3}{-}}

\newcommand{\fact}[2]{\textsf{\small [#1$\,\vert\,$\uppercase{{\scriptsize#2}}]}}
\newcommand{\entity}[1]{\textsf{\small #1}}
\newcommand{\attribute}[1]{\textsf{\small\uppercase{{\scriptsize#1}}}}

\journal{Information and Software Technology}

\begin{document}

\begin{frontmatter}

\title{A Bayesian Network Approach to Assess and Predict Software Quality Using Activity-Based
 Quality Models}

\author{Stefan Wagner}
\address{
Technische Universit\"at M\"unchen, Fakult\"at f\"ur Informatik\\
Boltzmannstr.~3, 85748 Garching, Germany\\
E-mail: wagnerst@in.tum.de\\
Tel.~+49 89 289 17334, Fax +49 89 289 17307
}

\begin{abstract}
\textbf{Context:}
Software quality is a complex concept. 
Therefore, assessing and predicting it is still
challenging in practice as well as in research. Activity-based quality models
break down this complex concept into concrete definitions, more precisely
facts about the system, process, and environment as well as their impact on
activities performed on and with the system. However, these models lack
an operationalisation that would allow them to be used in assessment and 
prediction of quality. Bayesian networks have been
shown to be a viable means for this task incorporating
variables with uncertainty. 
\textbf{Objective:}
The qualitative knowledge contained in activity-based quality models
are an abundant basis for building Bayesian networks for quality assessment.
This paper describes a four-step approach for deriving systematically a
Bayesian network from a assessment goal and a quality model.
\textbf{Method:}
The four steps of the approach are explained in detail and with
running examples.
Furthermore, an initial evaluation is performed, in which data from 
NASA projects and an open source system is obtained.
The approach is applied to this data and its applicability is analysed.
\textbf{Results:}
The approach is applicable to the data from the NASA projects and
the open source system. However, the predictive results vary depending
on the availability and quality of the data, especially the underlying
general distributions.
\textbf{Conclusion:} The approach is viable in a realistic context but
needs further investigation in case studies in order to analyse
its predictive validity.

\end{abstract}


\end{frontmatter}

\section{Introduction\label{sec:intro}}

Despite the importance of software quality, its management
is still an immature discipline in software engineering research and practice.
Research work has gone in many directions and produced a variety of
useful results. However, there is still no commonly agreed way for
quality management. The practice varies strongly from a concentration on
testing to a large-scale quality management process. 

 One main problem is that many of the tools and methods in quality assurance
 and management work isolated. For example, development teams usually
 perform peer reviews whose results are often not set into relation to results
 of integration or system tests performed by the quality assurance team.
 Hence, quality is tackled on many levels without a combined
 strategy \cite{wagner:wosq07}. What is missing is a clear integration of these
 single efforts. One prerequisite for such an integration is a quality management
 sub-process in the overall development process. The process defines the
 roles, activities, and artefacts and how they work together. Hence, the maturity
 level of the organisation's processes plays an important role. A second
 prerequisite is a clear quality model that specifies the quality of the software to
 be developed.

\subsection{Problem}
Current quality models such as the ISO 9126 \cite{iso9126-1:2001} have
widely acknowledged problems \cite{kitchenham96,deissenb:icsm07,wagner:esem09}.
Especially as a basis for assessment and prediction, the defined ``-ilities''
are too abstract. A clear transition to measurements is therefore difficult in practice.
Hence, quantitative quality assessment and prediction is usually done
without direct use of such a quality model. This, in turn, leads to isolated
solutions in quality management.

\subsection{Contribution} 

We use the previously proposed activity-based quality models (ABQM)
\cite{deissenb:icsm07} as a basis
for quality assessment and prediction. They provide a clear structure of
quality and detailed information about quality-influences. Activity-based
quality models have proven useful in practice to structure quality and to
generate corresponding guidelines and checklists. In this paper, we
add a systematic transition from ABQMs to Bayesian networks
in order to enhance their assessment and prediction capabilities.  A four-step approach
is defined that generates a Bayesian network using an activity-based
quality model and an assessment or prediction goal. The approach is
demonstrated in several examples.

\subsection{Outline}

We first motivate and introduce activity-based quality models in Section 
\ref{sec:quality_models}. In Section \ref{sec:approach} the four-step
approach for systematically constructing a Bayesian network from an
activity-based quality model is proposed. The approach is then
demonstrated in an initial evaluation in Section \ref{sec:proof} using
publicly available data from NASA projects and measured data from the
open source system \emph{Tomcat}. Related work is discussed
in Section \ref{sec:related} and final conclusions are given in
Section \ref{sec:conclusions}.

\section{Activity-Based Quality Models}
\label{sec:quality_models}

Quality models describe in a structured way the meaning of a software's 
quality. We introduce the use of general quality models and how the
modelling of activities and facts helps to define quality more precisely.

\subsection{Software Quality Models}

Not only the functionality but also the quality of a software
system needs to be specified in order to control it.
A quality model describes what is meant by \emph{quality} and refines
this concept in a structured way. In practice, this is often merely a
metric such as \emph{number of defects} or high-level descriptions as
given by the ISO 9126 \cite{iso9126-1:2001}.

In general,
there are two main uses of quality models in a software project: 

\begin{itemize}
\item As a basis for defining quality requirements 
\item For assigning quality assurance techniques and measurements to quality requirements
\end{itemize}

For the first use,
requirements engineers commonly constrain well-known quality attributes (reliability,
maintainability,
etc.) as defined in a quality model. In practice, they often reduce this
to simple statements such as ``The system shall be easily maintainable.''
The second use is often not explicitly considered. However, quality engineers
nearly always measure certain metrics 
such as the number of faults in the system 
detected by inspections and testing. The relationship between these
measures and quality attributes remains unclear. The reason lies in the
lack of practical means to define metrics for these high-level
quality attributes. Hence, quality models need more structure and detail 
to integrate them closely in the development process.

\subsection{Facts and Activities}

We proposed to use activity-based quality models \cite{deissenb:icsm07} in order
to address the shortcomings of existing quality models. 
The idea is to avoid to use high-level ``-ilities'' for defining quality and instead to break
it down into detailed facts and their influence on activities performed on and 
with the system.  In addition to information
 about the characteristics of a system,
the model contains further important facts about the process, the team and the
environment and  their respective influence on, for example, maintenance
activities such as \entity{Code Reading}, \entity{Modification}, or
\entity{Test}.
For example, redundant methods in the source code, also called
clones, exhibit a negative influence on modifications of the system, because
changes to clones have to be performed in several
places in the source code. Concrete models have been built for 
maintainability \cite{deissenb:icsm07}, usability \cite{winter:interactive07},
and security \cite{wagner:qaw09}.

For ABQMs, an explicit meta-model (also called structure model) was defined in order
to characterise the quality model elements and their relationships. Four
elements of the meta-model are most important: \entity{Entity}, 
\attribute{Attribute}, \entity{Impact} and \entity{Activity}.  An \entity{Entity}
can be any thing, animate or inanimate, that can have an influence on software
quality, e.g., the source code of a method or the involved testers. These 
entities are characterised by attributes such as \attribute{Structuredness} 
or \attribute{Conformity}. The combination of an entity and an attribute is
called a \emph{fact}. We use the notation \fact{Entity}{Attribute} for a
fact. For the example of code clones, we write \fact{Method}{Redundancy} to
denote methods that are redundant.
These facts are assessable
either by automatic measurement or by manual review. If possible, we define
applicable measures for the facts inside the ABQM.

An influence of a fact is specified by an \entity{Impact}. We concentrate
on the influences on activities, i.e., anything that
is done with the system. For example, \entity{Maintenance} or \entity{Use}
are high-level activities. The impact
on an \entity{Activity} can be positive or negative. An activity might 
have an influence on a fact that also has an impact on it. These cyclic situations
are not considered in the meta-model.

We complete the code clone example by adding the impact on 
\entity{Modification}: \impactn{Method}{Redundancy}{Modification}.
This means that if a system
entity \entity{Method} exhibits the attribute \attribute{Redundancy} it will
have a negative impact on the \entity{Modification} activity, i.e., changing
the method. 
A further example is the following tuple that
describes consistent identifiers:
\impactp{Identifier}{Consistency}{Modification}.
Its meaning is that identifiers that can be shown to be consistent have a positive
influence on the modification activity of the maintainer of the system. In the
model itself, we document more information such as textual descriptions,
sources, and assessment descriptions. However, the short notation captures
the essential relationships.

\begin{figure}[htb]
\centering%
\includegraphics[width=.7\textwidth]{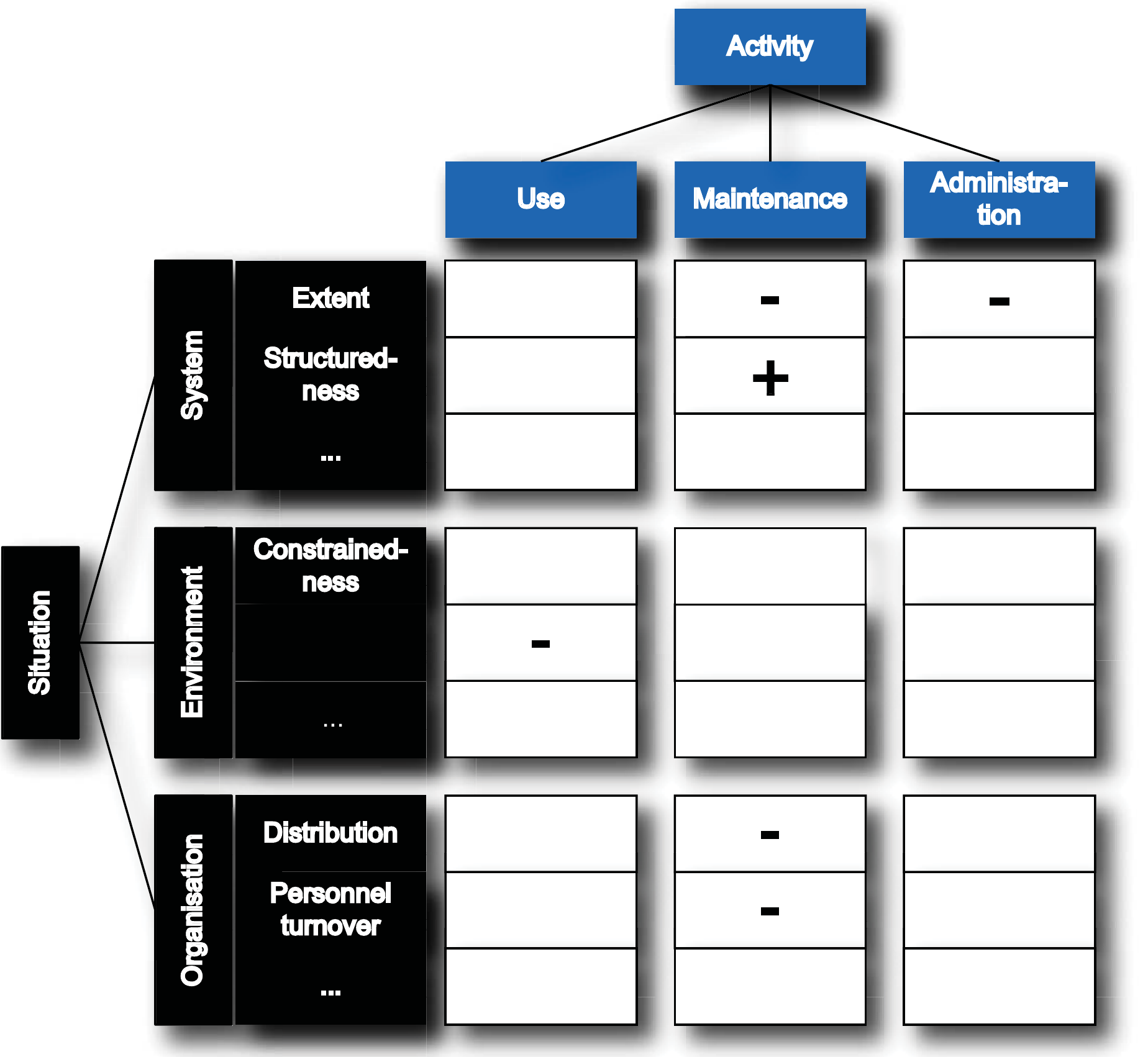}
\caption{High-level view on an activity-based quality model as a matrix}
\label{fig:quality_matrix}
\end{figure}

The model does not only contain the impacts of facts on activities but
also the relationships among these. Facts as well as activities are
organised in hierarchies. A top-level activity \entity{Activity} has
sub-activities such as \entity{Use}, \entity{Maintenance}, or
\entity{Administration}. These examples are depicted in Figure
\ref{fig:quality_matrix}. In realistic quality models they are further
refined. For example, maintenance can have sub-activities such as 
\entity{Code Reading} and \entity{Modification}. These activities can
be derived from existing standards (e.g., \cite{iso14764}) or the activities
defined in process models.

Because facts are a composite of an entity and an attribute,
the organisation in a hierarchy is straightforward. Hierarchical relationships
between entities do usually already exist. The top-level in Figure 
\ref{fig:quality_matrix} is the \entity{Situation} of the
software development project, which denotes the root for all entities of the
system as well as entities from its environment. In the example, it contains the \entity{System},
the system's \entity{Environment} and the development \entity{Organisation}. Again, these
entities need to be further refined. For example, the system could consist of
the source code as well as the executable.
All of these entities can be described with attributes, e.g., the
\attribute{Structuredness} of the \entity{System}. 
In principle, there could be more complex relationships instead of hierarchies,
but modelling and finding information tends to be easier if such complex
relationships are mapped to hierarchies.

The two hierarchies, the fact tree and the
activity tree, together with the impacts of the facts on the activities can
then be visualised using a matrix as in Figure \ref{fig:quality_matrix}. The 
fact tree is shown
on the left, the activity tree on the top. The impacts are depicted by entries
in the matrix where a ``+'' denotes a positive and a ``--''
a negative impact.

The associations between facts in the fact tree can have two different meanings.
Either an entity is a part or a kind of its super-entity. Along the 
inheritance associations, parts and attributes are inherited. Hence, it allows
a more compact description and prevents omissions in the model. For example,
naming conventions are valid for all identifiers no matter whether
they are class names, file names, or variable names.

Having defined all these entries in the ABQM, we can specify which
activities we want to support and which influencing facts need to be
analysed. In terms of the above example, if we want to support the activity
\entity{Modification}, we know that we need to inspect the identifiers for
their consistency. 

Another way of looking at ABQMs is as GQM patterns \cite{basili94,lindvall05}. The
activity defines the goal and the facts are questions for that goal that are measured
by certain metrics in a defined assessment. In the example, the goal is to evaluate
the modification activity that is analysed by asking the question ``How consistent
are the identifiers?'', which is asked in an assessment.

There exists a prototype tool to define this kind of large and detailed
quality models \cite{deissenb:icsm07}. Besides the easier creation and
modification of the model, this has also the advantage that we can automate
certain quality assurance tasks. For example, by using the tool we can
automatically generate customised review guidelines for specific views.

%%%%%%%%%%%%%%%%%%%%%%%%%%%%%%%%%
\section{Assessment and Prediction Approach}
\label{sec:approach}

Although activity-based quality models have proven to be useful in
practice, there is no systematic measurement approach for them.
Hence, there are no quantitative assessments and
predictions possible so far. We propose an approach that can be used
for systematically deriving assessment and prediction models from an
activity-based quality model.

\subsection{Aim and Basic Idea}

The general aim of the approach is to provide quality managers with
a systematic method to derive assessments and predictions from an
activity-based quality model. In the ABQM, there are definitions
of what quality means with respect to different situations, artefacts, and considered
activities. At present, we give a textual description in the quality model
that specifies how a fact could be assessed. For example, the fact described by
\fact{Method}{Redundancy}
contains the following assessment description: ``This fact can be assessed
manually or semi-automatically. For the automatic assessments there are
tools such as ConQAT or CCFinder to detect redundant parts of the
source code.'' This information is useful for quality assurance planning
but cannot directly be used for an overall assessment, let alone prediction.

Moreover, as the basic principle of activity-based quality models is that the
most important question in quality is how well activities can be performed on
and with the system, not only facts but also activities should be assessed and predicted
from the knowledge of facts and impacts. At present, activity-based quality models
only make the qualitative statement whether an impact is positive or negative. This is suitable
for rough assessments only. More comprehensive and precise assessments
of the current state and prediction of future states need a more sophisticated approach.
It has to systematically help using the given relationships and enriching
them with quantitative information. In terms of measurement scales, we move
from an ordinal scale to an interval scale or higher.

As most facts and especially the relationships between facts and activities
have an associated uncertainty, statistical methods are necessary. The two
major reasons are 

\begin{enumerate}
\item that we
cannot determine the exact relationship but can derive an uncertain range and
\item measured values can be uncertain, e.g., values from expert opinion. 
\end{enumerate}

Moreover,
the statistical method needs to be able to directly model the dependencies of
different factors from the quality model. We identified Bayesian networks as
most suitable for that task.

\subsection{Bayesian Networks}

\begin{figure}[htb]
\centering
\includegraphics[width=0.5\textwidth]{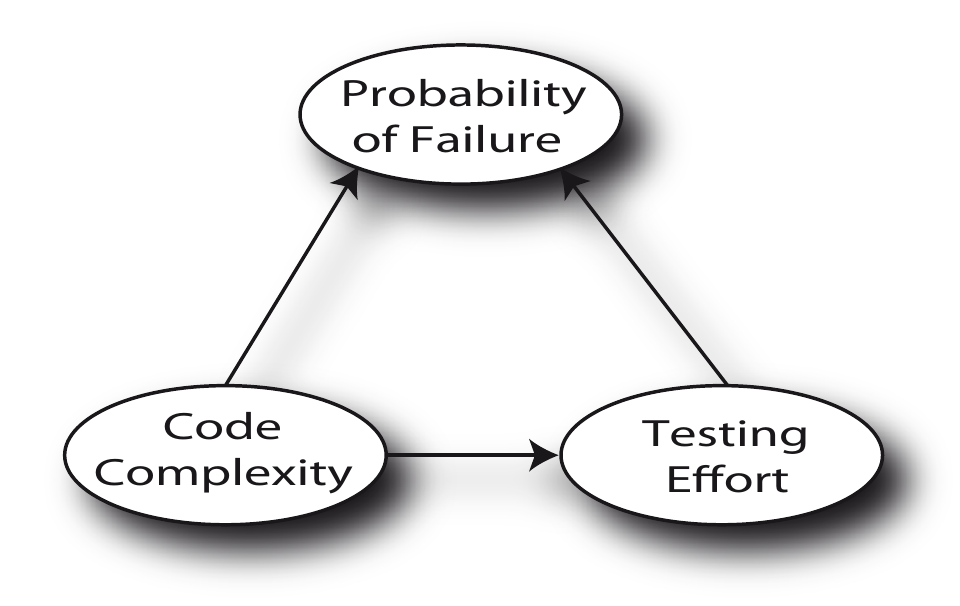}
\caption{A simple example Bayesian network}
\label{fig:bbn_example}
\end{figure}

Bayesian networks, also known as Bayesian belief nets or belief
networks, are a modelling technique for causal relationship based on
Bayesian inference. They are represented as a directed acyclic graph
with nodes for uncertain variables and edges for directed relationships
between the variables. This graph models all the relationships abstractly.
A hypothetical example with the 3 variables \emph{Code Complexity}, 
\emph{Testing Effort}, and \emph{Number of Field Failures} is given
in Figure~\ref{fig:bbn_example}. The code complexity influences the testing
effort and the number of field failures of a software. The testing effort also
impacts the number of failures.

For each node or variable there is a corresponding \emph{node probability 
table} (NPT). These tables define the relationships and the uncertainty of
these variables. The
variables are usually discrete with a fixed number of states. For each
state, it gives the probability that the variable is in this state. If
there are parent nodes, i.e., a node that influences the current node,
it defines these probabilities in dependence on the states of the 
parents. An example for the variable \emph{Number of Field Failures} is shown
in Table~\ref{tab:example_npt}. 
It specifies all combinations, e.g., that the \emph{Number of Field Failures} is with a probability of 60\% in 
the state ``$>100$'' if both parents are in the state \emph{low}, and 
with 40\% in ``$>100$'' if the testing effort is \emph{high} and the code complexity is \emph{low}.

\begin{table}[htb]
  \caption{An example NPT for the variable \emph{Number of Field Failures} with two states and two parents
          \label{tab:example_npt}}
\begin{center}
\begin{tabular}{lrrrrrr}
\hline
Testing Effort & & low & & & high & \\
\hline
Code Complexity & low & med & high & low & med & high\\
\hline
$<100$ & 0.4 & 0.3 & 0.2 & 0.6 & 0.55 & 0.5\\
$>100$ & 0.6 & 0.7 & 0.8 & 0.4 & 0.45 & 0.5\\
\hline
\end{tabular}
\end{center}
\end{table}

The process of building a Bayesian network contains the identification
of important variables that shall be modelled, representing them as nodes,
constructing the topology and specifying the NPTs. Each of these steps
is important and non-trivial. First, the identification of \emph{important}
variables includes the assumption that the model builder can decide on
some basis what is important. In many cases, this is not clear beforehand.
One possibility is to include many variables and use sensitivity analysis to
remove insignificant ones. A model of the complete situation is usually
not feasible, because the network becomes too complex, very elaborate
to build, and most often there is no knowledge available about several
variables.

Second, the creation of the topology utilises the assumption that the model
builder can decide on the dependence and independence of the
identified variables. In the process of building the Bayesian network,
especially for independence assumptions (i.e., missing edges in the
graph), detailed justifications should be given. Third, the problem of
constructing NPTs is widely acknowledged in the literature \cite{fenton:tr07}. 
A major part of this problem is that it involves
defining quantitative relationships between variables. There are various
possible methods for this quantification such as a probability wheel or
regression from empirically collected data. All methods have their pros and
cons.

It is important to note that each of these steps is important and errors
in each of these steps can have a large effect on the outcome. Bayesian
networks and the corresponding tool support make it easy to build models
and get quantitative results. However, one needs to be aware that many
assumptions are embedded in a Bayesian network that need to be validated
before it can be trusted.

\subsection{Four Steps for Network Building}

We propose a four-step approach for building a Bayesian network
as assessment and prediction model derived systematically from an
activity-based quality model. The resulting Bayesian network contains three
types of nodes:
\begin{itemize}
\item \emph{Activity nodes} that represent activities from the quality model
\item \emph{Fact nodes} that represent facts from the quality model
\item \emph{Indicator nodes} that represent metrics for activities or facts
\end{itemize}
We need four steps to derive these nodes from the information of the
ABQM. First, we identify the relevant activities with indicators
based on the assessment or prediction goal. Second, influences by
sub-activities and facts are identified. This step is repeated recursively
for sub-activities. The resulting facts together with their impacts are modelled.
Third, suitable indicators for the facts are added. Fourth, the node probability
tables (NPT) are defined to reflect the quantitative relationships. Having that,
the Bayesian network can be used for simulation by setting values for any
of the nodes. However, we first describe the four steps in more detail.

\paragraph{Step 1}
The first step is a goal-based derivation of relevant activities and their indicators.
We use GQM \cite{basili94} to structure that derivation. We first define the
assessment or prediction \emph{goal}, for example, optimal maintenance
planning or optimisation of security assurance. The identification and resolution
of conflicting goals is out of scope of this method. The goal leads to relevant
activities, such as \emph{maintenance} or \emph{attack}. This is refined by
stating \emph{questions} that need to be answered to reach that goal. For example, 
how high will be the maintenance effort over the next year or how often will there
be a successful attack in the next year? Finally, we derive \emph{metrics} or
indicators that allow a measurement to answer the question. In the examples,
it can be the average change effort or the number of harmful attacks over the next year.

\paragraph{Step 2}
In the second step, we use the quality model to identify the other factors that
are related to the identified activities. There are two possibilities: 

\begin{enumerate}
\item There are sub-activities of the identified activities.
\item There are impacts from facts to the identified activities. 
\end{enumerate}

\begin{figure*}[htb]
\centering
\includegraphics[width=\textwidth]{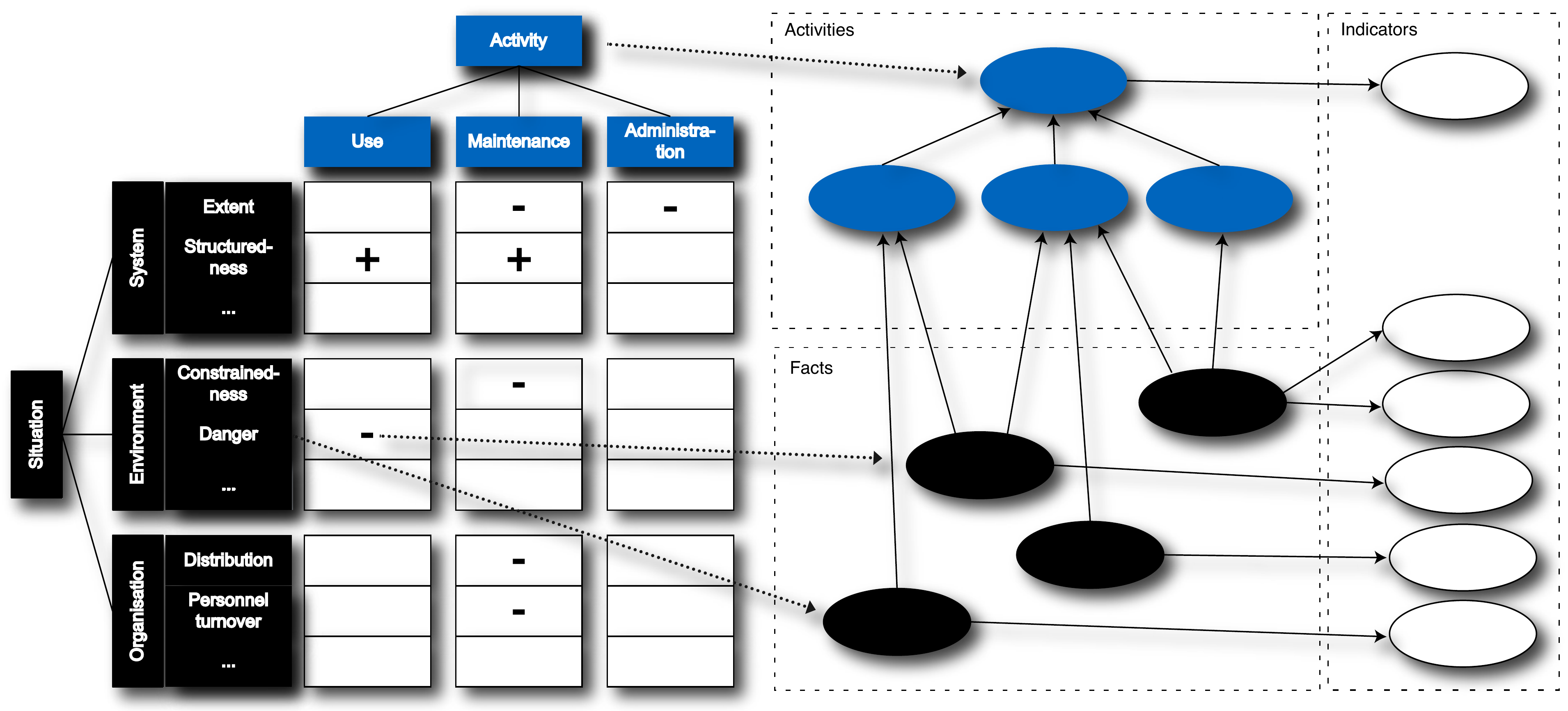}
\caption{Mapping of quality model elements to Bayesian network elements.
               The dashed arrows indicate examples for mappings.}
\label{fig:qm2bbn}
\end{figure*}

We repeat this recursively for the sub-activities until
all facts are collected that have an impact on the activities sub-tree below the
identified activity. For each activity we immediately see the impacts and hence
the corresponding facts. All activities and facts identified this way are modelled
as nodes in the Bayesian network. We add edges from sub-activities to
super-activities and from facts to activities on which they have an impact.
Figure \ref{fig:qm2bbn} gives an abstract overview of that mapping from the
quality model to the Bayesian network.

\paragraph{Step 3}
In the third step, we add additional nodes as indicators for each fact and
activity node that we want a measurement for. In the first step, we defined 
the indicator for our relevant activity. Hence, we can add additional indicators
for sub-activities if needed. In any case, there need to be at least one indicator
for each fact that is modelled. There might be a precise description in the
quality model already. Otherwise, we need to derive our own metric or use an
existing one from the literature. The indicator does not have to be measured
automatically, but manual reviews can also be included in the assessment.
The edges are directed from the activity and fact nodes to the indicators, i.e., the
indicators are dependent on the facts and activities as an indicator is only an expression
of the underlying factor it describes.

A main advantage of using an ABQM as a basis for the Bayesian network
is that it prescribes its topology. One of the prime points of such quality models
is to qualitatively describe the relationships between different factors that
are relevant for software quality. We rely on that and assume that all
dependencies have been modelled and that all other factors are independent.
On the one hand, this constrains the validity of the results of the Bayesian network by the
validity of the ABQM. On the other hand, it frees the network builder from
reasoning about independence and dependence.

\paragraph{Step 4}
Finally, the fourth step enriches the Bayesian network with quantitative information.
This includes defining node states as well as filling the NPT for each node. The
activity and fact nodes are usually modelled as \emph{ranked} nodes, i.e., in an
ordinal scale. The most common example is the scale containing \emph{low},
\emph{medium}, and \emph{high}.
This has advantages in evaluation and aggregation. The evaluation
is easier as not precise numbers have to be determined but coarse-grained levels.
These levels actually reflect much more the high uncertainty in the data. In
aggregating over nodes (up the hierarchy in the activity tree) coarse-grained ranked
data is also more simple to handle. It is easier to define an aggregation specification 
for only a few levels in an ordinal scale than for continuous data. 

To define the NPT,
we use an approach proposed by Fenton, Neil and Galan Caballero \cite{fenton07a}.
The basic idea is to formalise the behaviour observed with experts that have to
estimate NPTs. They usually estimate the central tendency or some extreme values
based on the influencing nodes. The remaining cells of the table are then filled
accordingly. This is similar to linear regression where a Normal distribution is
used to model the uncertainty. We use the doubly truncated Normal distribution
(TNormal)
that is only defined in the $[0,1]$ region. It allows to model a variety of shapes
depending on the mean and variation. For example, it renders it possible to model the NPT
of a node by a weighted mean over the influencing nodes.

The node states of indicator nodes depend on the scale of the indicator used.
This often will be continuous or discrete interval states such as lines of codes
in intervals of a hundred or a thousand. The NPTs of the indicator nodes
are then defined using either common industry distributions or information from
company-internal measurements. For example, typical LOC distributions can be
accumulated over time. The influence of the activity or fact node
it belongs to can be modelled in at least two ways: 

\begin{itemize}
\item partitioned expressions
\item arithmetic expressions
\end{itemize}

The latter describes a direct arithmetical
relationship from the level in the activity or fact node to the indicator. Using
a partitioned expression, the additional uncertainty can be expressed by
defining probability distributions for each level of the node.

\subsection{Usage of the Bayesian Network}

A major feature of Bayesian networks is their capability to simulate different
scenarios. Having built the Bayesian network based on the ABQM, 
we can ask ``what if?'' questions. These questions are formulated in
scenarios that can be simulated and compared. A scenario involves adding
additional information to the model, more specifically, adding an observation for
a node. This way, the uncertainty is removed and the consequences for the other
nodes can be calculated. In a Bayesian network it is possible to do
forward as well as backward inference, i.e., information can be added to any node
and the effect is calculated in any direction of the graph.

A first straightforward scenario is to add the currently measured values for the
fact indicator nodes. This will drive the calculation up to the activities and the
activity indicator node. The activity indicator node then shows the probability distribution
for its value, i.e., the value of the activity. Afterwards changes can be made to the
fact indicator nodes in other scenarios to reflect possible changes and their effect
on the activity can be predicted. A further interesting scenario is to set a desired
value for the activity indicator and let the network calculate the most probable
explanation in the fact indicators. It provides the values that should be reached
in order to fulfil the goal.

%%%%%%%%%%%%%%%%%%%%%%%%%%%%%%%
\section{Initial Evaluation}
\label{sec:proof}

The approach presented above can be used in various contexts 
to answer assessment and prediction questions. In this section, we provide 
an initial evaluation by applying the approach to several publicly available data
sets as well as automatically collectable measures for an open source system.

\subsection{Goal}

We provide an exemplary application of the approach
based on a small extract of our quality model for maintainability for which there
is public data available, and on our model for security for which we
collect measures from an open source system. The examples 
demonstrate the basic principles of the approach
on real data sets. This way, we analyse whether the approach is
feasible in an almost realistic setting. The predictive validity in the examples
can give an indication of the usefulness of the approach, because we
expect an improvement over industry average values at least.

At present, we cannot provide a complete validation of the assessment and prediction
approach because this would involve measuring a large number of facts from
the quality model in order to take full advantage from the knowledge contained
in it. This data is not available in public data sets. Measuring this at a company
will need time and effort. Only then a sensible analysis of the predictive validity
and comparisons with other prediction models are possible.

\subsection{Maintainability Cases}
\label{sec:maintainability_cases}

The first part of the examples deals with the analysis of maintainability.
We follow the IEEE and define maintainability as follows: ``The ease with which
a software system or component can be modified to correct faults, improve 
performance or other attributes, or adapt to a changed environment.''
\cite{ieee610} 

\subsubsection{Context}

The 4 systems under analysis for maintainability were developed by NASA in the projects
CM1, KC1, KC3, and KC4  for which the data has been publicised \cite{nasa-mdp}. The 
characteristics of these projects are summarised in Table~\ref{tab:characteristics}.
Various metrics, including the McCabe and Halstead metrics, have been collected in these
projects. The projects were selected for this analysis because their data is openly available and effort
data per defect was collected. Hence, the selection criterion was opportunistic. Nevertheless,
the diversity of sizes, functionality, and programming language in the set of analysed systems
is sufficient to generalise the experiences to other systems.

\begin{table}[htp]
\caption{Characteristics of the projects used in the maintainability cases}
\begin{center}
\begin{tabular}{lrll}
\hline
Project & Size (LOC) & Lang. & Function\\
\hline
CM1 & 16,903 & C & Space craft instrument\\
KC1 & 42,963 & C++ & Ground system\\
KC3 & 7,749 & Java & Satellite data processing\\
KC4 & 25,436 & Perl & Subscription server\\
\hline
\end{tabular}
\end{center}
\label{tab:characteristics}
\end{table}%

As the quality model, we use the activity-based quality model for maintainability
from \cite{deissenb:icsm07}. It contains a complete activity tree for maintenance
as well as about 200 facts with an impact on those activities. The NASA
data sets do not contain data for all of these facts; hence, we select a small set
of facts and corresponding activities to predict maintainability. The choice was
therefore guided by the availability of the data. The actual maintainability can be
judged for these projects, because the effort for several changes has been documented.
We will use this as our surrogate measure.

For modelling the Bayesian network, we use the tool 
AgenaRisk\footnote{\url{http://www.agenarisk.com/}}. It provides
a complete tool environment for Bayesian networks including the usage of
expressions for describing the NPT and sensitivity analysis.

\subsubsection{Procedure}

The first step in our assessment and prediction approach is to identify the relevant
activities and corresponding indicators (metrics) in a GQM-like procedure. We
want to analyse maintainability and we assume that the quality manager is
interested in the question of how high the maintenance efforts will be on average in the
future. This information helps in planning the maintenance team. Hence, we
formulate the \textbf{goal} as ``Planning of future maintenance efforts''. We
can directly identify the activity \entity{Maintenance} in that goal. To further
operationalise that, we define the corresponding \textbf{question} as ``What
will be the maintenance effort per change request?''. This information, 
together with a prediction of the yearly number of change requests, would give
a good basis for maintenance planning. However, we concentrate only on the
question about the effort per change request. This leads us straightforwardly
to the \textbf{metric} ``average effort per change request''. 

\begin{figure}[htb]
\centering
\includegraphics[width=.8\textwidth]{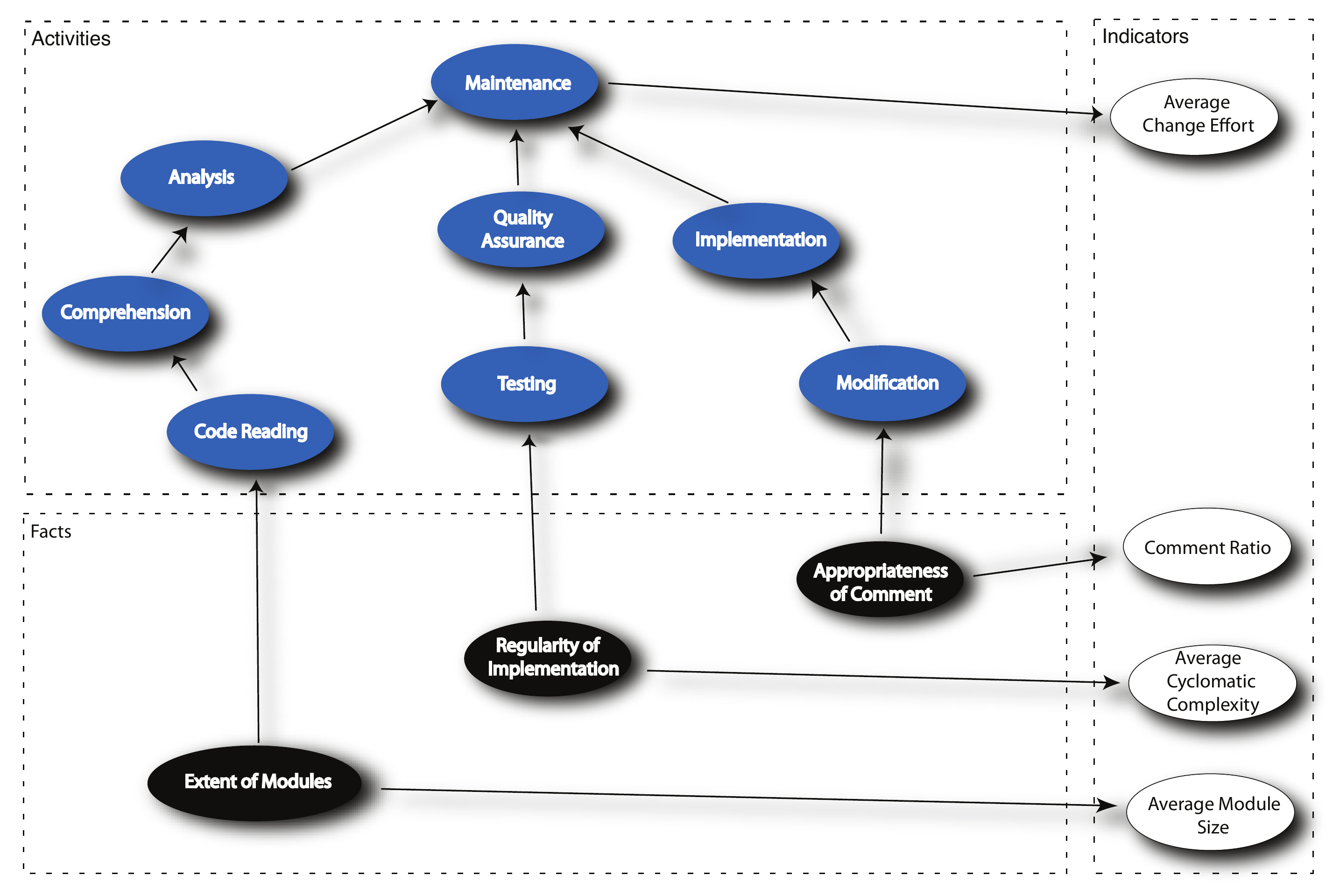}
\caption{The topology of the Bayesian network for the maintainability cases}
\label{fig:maintain_topology}
\end{figure}

In the second step, we start building the Bayesian network. We look
at the maintainability model and find no direct impacts on \entity{Maintenance}
that should be considered here. However, we find 10 sub-activities including the
following 3 that we analyse further: \entity{Quality Assurance}, \entity{Implementation}
and \entity{Analysis}.  All three have impacts from facts but having in
mind the available data, we ignore these and use the further sub-activities
\entity{Testing}, \entity{Modification} and \entity{Comprehension} with its child
\entity{Code Reading}. We create nodes for these activities and connect them with
edges corresponding to their hierarchy. They can be seen in Figure \ref{fig:maintain_topology}
in the box ``Activities''.

We only include 3 impacts on each of the lowest level activities chosen so far
because we can find corresponding data in the NASA data sets. These facts together
with their impacts are:
\begin{itemize}
\item
\impactn{Module}{Extent}{Code Reading}\\
The size of a module has an impact on reading the code of the module. In essence,
the larger the module, the longer it takes to read it.

\item
\impactp{Implementation}{Regularity}{Testing}\\
An implementation is regular if it does not use unnecessarily nested branches. This
complex structure would render coverage by tests more difficult.
\item
\impactp{Comment}{Appropriateness}{Modification}\\
Comments need to appropriately describe the code it is associated with.
\end{itemize}

We add these facts as nodes in the Bayesian network into the box ``Facts'' 
in Figure \ref{fig:maintain_topology}. The impacts are included as the arrows
from the facts to the activities. There are further impacts conceivable, but we restrain
them to the impacts available in the used maintainability model \cite{deissenb:icsm07}.

The indicators that are identified in the fourth step of the approach are taken from
the available data of the data set. We only use one indicator per fact although we are
aware that each fact has more aspects that should be covered. For the \emph{Extent
of Modules}, we use the indicator \emph{Average module size} given in LOC.
The \emph{Regularity of the Implementation} is indicated by the \emph{Average
Cyclomatic Complexity}. This is not a particularly good indicator as it only gives a
number for the decision points in the implementation. In contrast, a manual review could far better
decide whether the implementation is regular. However, we have no access to review
results. A similar reasoning holds for the indicator \emph{Comment Ratio} for the fact
\emph{Appropriateness of Comments}. The proportion of the comments in relation to
the other code is only of minor importance in comparison with the semantic appropriateness.
However, we do not have access to such a semantic judgement. The indicators can
be found in Figure \ref{fig:maintain_topology} on the right-hand side in the box ``Indicators''.

A difficult problem in general with Bayesian networks is the definition of the node
probability tables (NPT)  \cite{fenton:tr07}. Network builders can use various methods
to define these NPTs, which we do not describe here in detail. For the approach, we can simplify
the problem to 2 cases: 

\begin{itemize}
\item Activity and fact NPTs
\item Indicator NPTs
\end{itemize}

The NPTs for the activities and facts can be assumed as uniformly distributed unless there
is additional knowledge. The impacts are only modelled as direct influence. Here, a
special method for ranked nodes \cite{fenton07a} can be applied that simplifies the
task. For example, the node \emph{Maintenance} is modelled by the following distribution
as its NPT:
\begin{equation}
\mbox{TNormal}\left[\frac{\mbox{Analysis}+\mbox{Quality\ Assurance}+
\mbox{Implementation}}{3}, \frac{1}{3}, 0.001\right]
\end{equation}
The values of the three nodes \emph{Analysis}, \emph{Quality Assurance}, and
\emph{Implementation} are averaged with the implicit weight 1 each. The 1/3 gives
the normalisation and 0.001 is the variance.

For the indicator NPTs either 
empirically investigated distributions of the 
company or industry average distribution should be used. This forms the distribution
of the indicator values under uncertainty without any observations.
The indicator node \emph{Comment Ratio} is shown as an example in 
Table~\ref{tab:indicator_npt}. It gives the NPT as an partitioned expression
depending on the state of the parent node \emph{Appropriateness of Comments}.

\begin{table}[htp]
\caption{NPT of the indicator node \emph{Comment Ratio}}
\begin{center}
\begin{tabular}{llll}
\hline
Appropriateness of Comments & Expressions\\
\hline
Low & TNormal[0.01,1,0.03]\\
Medium & TNormal[0.1,1,0.05]\\
High & TNormal[0.25,1,0.1]\\
\hline
\end{tabular}
\end{center}
\label{tab:indicator_npt}
\end{table}%

For the average effort of a change, we refer to \cite{wagner:isese06} that gave
a mean defect removal cost of 27.4 person-hours with minimum 3.9 and maximum
66.6. Although a change does not always have to be a defect removal, it is precise
enough for the initial evaluation. For the distributions of the other indicators, no published
distributions were available. Hence, our own expert opinion was used as a basis.

\subsubsection{Results and Discussion}

\begin{figure}[htb]
\centering
\includegraphics[width=\textwidth]{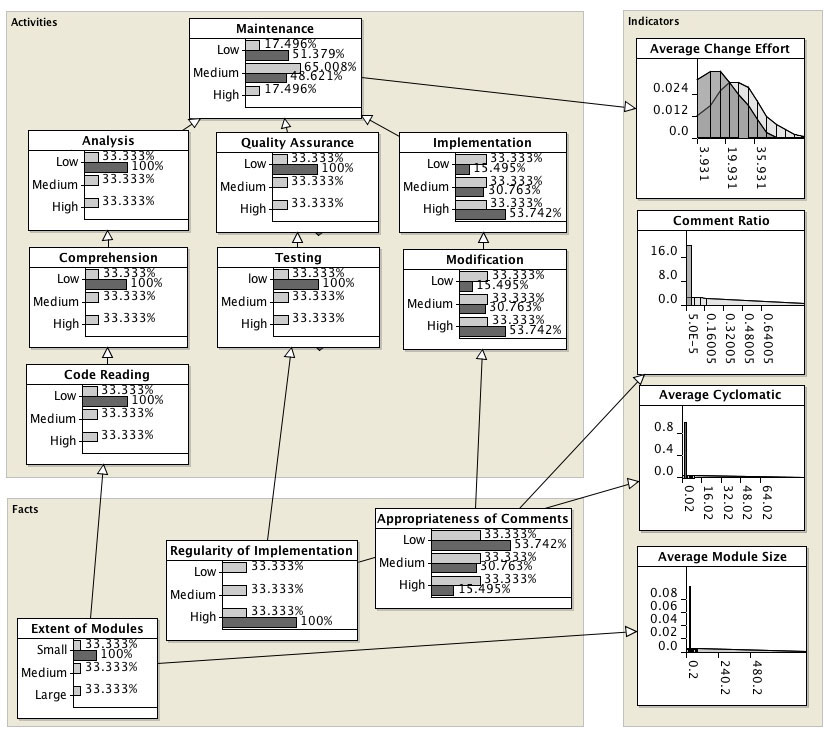}
\caption{The Bayesian network for KC1 in AgenaRisk. The standard scenario is
shown in light grey, the KC1 scenario in dark grey.}
\label{fig:screenshot}
\end{figure}

Figure \ref{fig:screenshot} shows a screenshot from AgenaRisk as an example. The light-grey values describe the
general scenario in which no observations have been made. The
dark-grey values are for the KC1 scenario in which for the three fact
indicators the real measured values are set as observations. Table~\ref{tab:maintainability_results}
summarises the used observation data and the results from the Bayesian network
for all 4 maintainability cases. It shows the observed values for the comment ratio,
the average cyclomatic complexity, and the average module size in LOC.
These were fixed in their corresponding scenarios. This has effects on the rating of the facts and in
turn on the activities. The most interesting value, however, is the average change
effort. In the general scenario, this variable has a mean value of 27 with a standard
deviation of 12.1. The table gives the predicted average change effort in person hours
together with the standard deviation as well as the actually observed average change effort.

\begin{table}[htp]
\caption{Observations and results for the maintainability cases}
\begin{center}
\begin{tabular}{lrrrr}
\hline
 & CM1 & KC1 & KC3 & KC4\\
\hline
Comment ratio & 0.25 & 0.02 & 0.08 & 0.00\\
Avg.\ cyclomatic complexity & 5.18 & 2.84 & 3.45 & 10.05\\
Avg.\ module size & 33.47 & 20.39 & 16.92 & 203.49\\
\hline
Predicted avg.\ change effort & 15.9 & 19.4 & 19.2 & 36.1\\
Standard deviation & 8.5 & 9.8 & 9.8 & 12.1\\ 
Observed avg.\ change effort & 6.0 & 21.7 & 24.8 & 12.1\\
\hline
\end{tabular}
\end{center}
\label{tab:maintainability_results}
\end{table}%

The 4 cases show 3 different types of results. For KC1 and KC3, the
predictions are very close to the observed values with a difference of
2.3 and 5.6 person-hours per change, respectively. For both predictions,
the observed values are well inside the standard deviation. Hence, in
these cases, the predictions are reasonably accurate.

For CM1, the additional information about the measured values shifts the
distribution to the left. The mean decreases to 15.9 with a standard deviation of 8.5. This
is a deviation of 9.9 from the observed value. However, it is much closer than the industry standard
of 27. Hence, it is an improvement in comparison to just using standard numbers.

In the case of KC4, the industry standard 27 \cite{wagner:isese06} would have been closer than the
prediction of 36.1 which constitutes a difference of 24 from the actual value.
This is not even inside the standard deviation. The prediction is far too pessimistic.
The reason for the large difference between the prediction and the real value
can be  three-fold. First, the effort distribution from \cite{wagner:isese06} might 
not be appropriate for the NASA environment. Second, it might also be the
case that the data in the NASA data sets do not use exactly the same measures as
in \cite{wagner:isese06}. The degree to which additional efforts for re-inspection
and re-testing are included could vary. Third, several more factors than the 3
considered might have an influence on the maintenance effort. The quality
model contains many explanations in terms of facts that should be investigated. 
Especially for KC4, there are probably other factors that decrease the change
effort in such a way. For example, the capabilities of the developers are known
to have a large effect on development productivity \cite{boehm00}.
However, we cannot include this as a fact in this case, because we have no
data available.

\subsection{Security Case}

The second part of the examples looks at security. We use
the following definition of Avizienis et al.:
``Security is a composite of the attributes of confidentiality, integrity, 
and availability, requiring the concurrent existence of 
1) availability for authorized actions only, 2) confidentiality, 
and 3) integrity with `improper' meaning `unauthorized.'{''}
\cite{avizienis04}

\subsubsection{Context}

The servlet container \emph{Tomcat}\footnote{\url{http://tomcat.apache.org/}} is
the reference implementation of the Java servlet and JSP specification
by the Apache Software Foundation. Its main goal is to deliver HTTP responses
dynamically assembled based on an HTTP request.  
For this, Tomcat is one of the most used systems in the world.
We employ version 6.0 of Tomcat, which implements the Java servlet specification
2.5 and the JSP specification 2.1. The initial version 6.0.0 was released in
December 2006. The current version 6.0.20 dates from May 2009. It consists of
over 300 KLOC of Java code and it is in real production on many sites.

A software application with such a fundamental role in Java web applications
has special responsibilities in the management of vulnerabilities.
Therefore, the Apache foundation
openly publishes the identified vulnerabilities. They can be found on the Tomcat
website\footnote{\url{http://tomcat.apache.org/security-6.html}}. 
This published list of vulnerabilities together with the availability of the
source code for further analyses is the reason for selecting
Tomcat for this example. Hence, the selection criterion was again
opportunistic.

The ABQM used for the security case is the security model as
described in \cite{wagner:qaw09}. This model uses a well-known
hierarchy of attacks as activities and a collection of open security
guidelines as facts. As modelling tool for the Bayesian network, 
we employ again AgenaRisk.

\subsubsection{Procedure}

We start with the first step of our assessment and prediction approach and
 identify the relevant activities and corresponding indicators. We
analyse security, in particular the risk of vulnerabilities in the system. 
The risk can be the basis for deciding whether further security improvements
need to be employed.  Therefore, the \textbf{goal} is ``Planning of further security
improvements''. For security improvements, attacks on the system need to
be confounded. Hence, the activity \entity{Attack} need to be analysed.
For the operationalisation, we define the \textbf{question} ``How many 
vulnerabilities are there in relation to the software size?''. For the security
improvement planning, it is
not only important how many vulnerabilities there are but also whether this
number is in a reasonable relation to the system size. It might be economically
inadvisable to invest in removing all vulnerabilities.  The corresponding \textbf{metric} 
\emph{vulnerability density} that measures the number of vulnerabilities by source
code size in KSLOC can be directly derived from the question. 

\begin{figure}[htbp]
\begin{center}
\includegraphics[width=.8\textwidth]{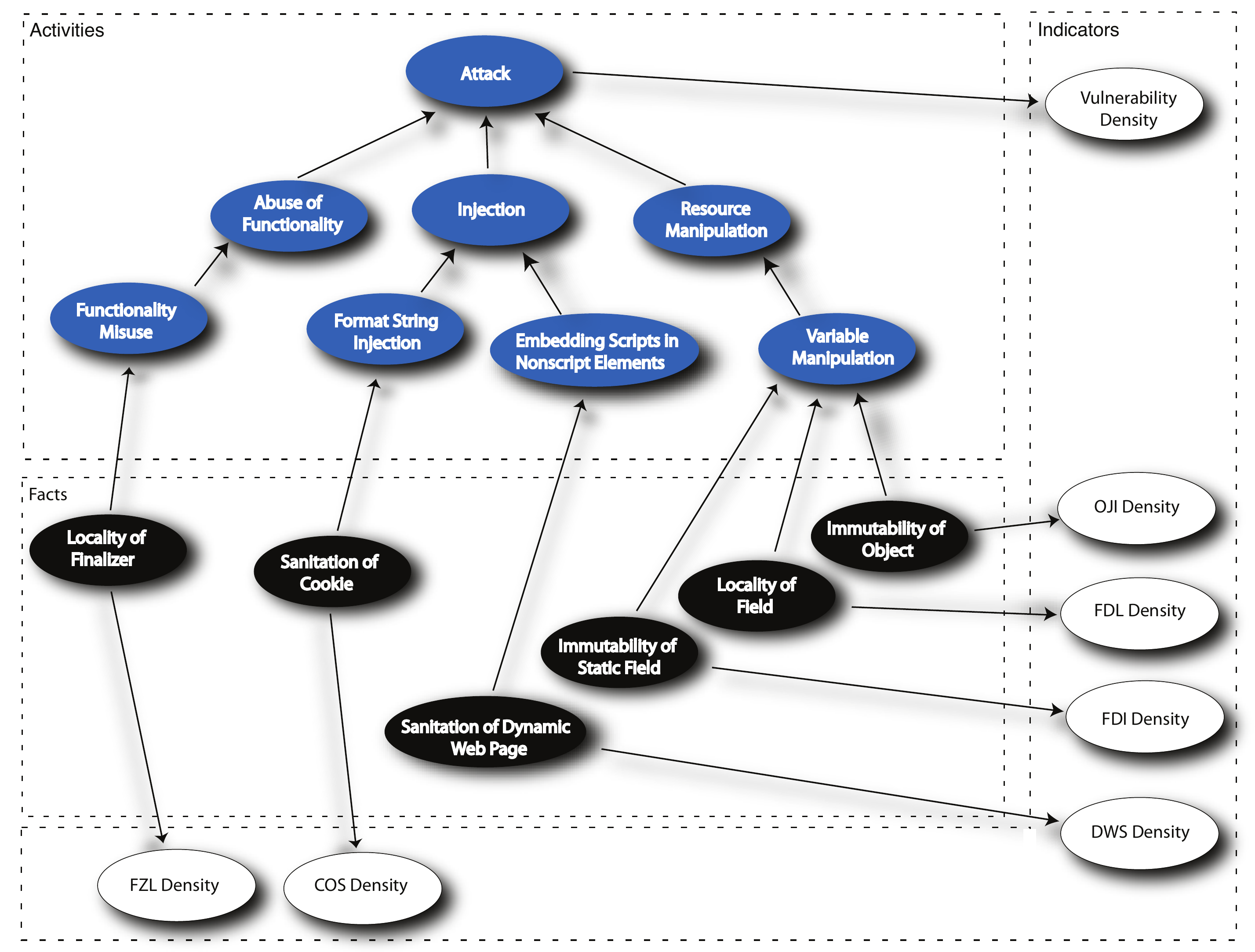}
\caption{The topology of the Bayesian network for the security case}
\label{fig:security_topology}
\end{center}
\end{figure}

In the second step of the prediction and assessment approach, we build the
Bayesian network (see Figure~\ref{fig:security_topology}). In the ABQM for security, there is the top-level activity \entity{Attack}
that we measure by the above derived \emph{vulnerability density}. There are no
direct impacts on this activity. Therefore, we break it down to \entity{Abuse of Functionality},
\entity{Injection}, and \entity{Resource Manipulation}. The former is further refined into 
\entity{Functionality Misuse}, the second into \entity{Format String Injection} and
\entity{Embedding Scripts in Non-Script Elements}, the latter into 
\entity{Variable Manipulation}. In the Bayesian network, nodes are created for these
activities w.r.t.\ the given hierarchy.

We can include 8 impacts on these activities. The impacts are chosen so that their
corresponding facts can be automatically measured by a bug pattern tool. We chose
the open source tool \emph{FindBugs}\footnote{\url{http://findbugs.sourceforge.net/}},
because it is widely used and well evaluated. The facts used together with their impacts are:
\begin{itemize}
\item
\impactn{Object}{Immutability}{Variable Manipulation}\\
When objects are not immutable the caller can change the contents of these 
objects which may have security implications. Hence, objects should be immutable
if possible. 

\item
\impactn{Field}{Locality}{Variable Manipulation}\\
A mutable static field could be changed by malicious code or by accident from another
package. The field could be made package protected and/or made final
to avoid this vulnerability. 

\item
\impactn{Field}{Immutability}{Variable Manipulation}\\
When a static field is mutable or references mutable objects, these could be changed
by malicious code. 

\item
\impactn{Finalizer}{Locality}{Functionality Misuse}\\
Malicious code could call the method \emph{finalize} if it is not declared protected
while the object is still used and external resources could be closed too early.

\item
\impactn{Cookie}{Sanitation}{Format String Injection}\\
If a web product does not properly protect assumed-immutable values from
cookies, this can lead to modification of critical data. 

\item
\impactn{Dynamic Web Page}{Sanitation}{Embedding Script in HTTP Headers}\\
\impactn{Dynamic Web Page}{Sanitation}{Embedding Scripts in HTTP Query Strings}\\
\impactn{Dynamic Web Page}{Sanitation}{XSS in Error Pages}\\
If a web application does not sufficiently sanitise the data it is using in output, arbitrary
content, including scripts, can be included by attackers.

\end{itemize}

In the fourth step of the approach, indicators are defined for all 8 impacts. The indicators
in this case are all derived from \emph{FindBugs} bug patterns that are summed if they
only measure different aspects of the same fact and set into relation to the size of the
system. Hence, we end up with a set of densities of detected bug pattern.
Table \ref{tab:rules} gives the used facts together with their associated bug patterns
and the corresponding name of the density of the sum of the findings for these patterns.
The final topology of the Bayesian net is shown in Figure~\ref{fig:security_topology}.

\begin{table}[htp]
\caption{The mapping of facts to FindBugs patterns and metrics}
\begin{center}
\begin{tabular}{cp{7cm}c}
\hline
Fact & FindBugs Pattern & Metric\\
\hline
\fact{Object}{Immutability} &
\begin{compactitem}
\vspace{-.5em}
\item May expose internal representation by returning reference to mutable object 
\item May expose internal representation by incorporating reference to mutable object
\vspace{-.9em}
\end{compactitem} &
OJI\\
\hline
\fact{Field}{Locality} & 
\begin{compactitem}
\vspace{-.5em}
\item Field isn't final and can't be protected from malicious code
\item Field should be moved out of an interface and made package protected
\item Field should be package protected
\item Field isn't final but should be
\vspace{-.9em}
\end{compactitem} &
FDL\\
\hline
\fact{Field}{Immutability} & 
\begin{compactitem}
\vspace{-.5em}
\item May expose internal static state by storing a mutable object into a static field
\item Field is a mutable array
\item Field is a mutable Hashtable
\vspace{-.9em}
\end{compactitem} &
FDI\\
\hline
\fact{Finalizer}{Locality} & Finalizer should be protected, not public & FZL\\
\hline
\fact{Cookie}{Sanitation} & HTTP cookie formed from untrusted input & COS\\
\hline
\fact{Dyn.~Web Page}{Sanitation} & Servlet reflected cross site scripting vulnerability & DWS\\
\hline
\end{tabular}
\end{center}
\label{tab:rules}
\end{table}%

There is little quantitative data published on vulnerabilities. Therefore, we use the
data on vulnerability densities from Alhazmi, Malaiya, and Ray \cite{alhazmi05} who
analysed several releases of Microsoft Windows and Red Hat Linux. On average, these
systems had a known vulnerability density of 0.0054 with a minimum of 0.0022 and a
maximum of 0.0112 vulnerabilities per KSLOC. There is no data available about what are average
densities or even occurrences of specific bug patterns in software systems. Hence, we make
the coarse assumption that we consider a density of 0.4 as normal for all defined indicators for
the facts. That means that 4 occurrences of each of these bug patterns in 10 KSLOC is normal.
We employed an exponential distribution as we assume that there is a high 
concentration in the lower areas of the distribution that decreases quickly in the higher areas.

\subsubsection{Results and Discussion}

Using the analysis tool ConQAT \cite{deissenb:softw08}, we found that
the whole Tomcat 6.0.0 source distribution contains 306,675 LOC of Java code,
which correspond to 151,509 SLOC in 1030 files. This is the size baseline for
all the following densities. The total number of published vulnerabilities of
Tomcat is 24. However, we only consider the vulnerabilities rated as
\emph{important} and \emph{moderate} because we assume that in other
systems other vulnerabilities would not be counted as such. With this
assumption the number is reduced to 11.

Table~\ref{tab:security_results} contains the summarised results. First,
the 6 indicators are given. These densities range from 0 to 1.14 occurrences
per KSLOC. In the Bayesian network, this led to the prediction of 0.006
vulnerabilities per KSLOC with a standard deviation of 0.003. Actually observed
in Tomcat over all 6.0 versions, we found a vulnerability density of 0.07
vulnerabilities per KSLOC.

\begin{table}[htp]
\caption{Observations and results for the security case. All densities are given
in occurrences per KSLOC.}
\begin{center}
\begin{tabular}{lc}
\hline
OJI density &  1.14\\  
FDL density & 1.63 \\ 
FDI density & 0.06 \\ 
FZL density & 0.03 \\ 
COS density & 0.00\\
DWS density & 0.00\\
\hline
Predicted vulnerability density &  0.006\\
Standard deviation &  0.003\\ 
Observed vulnerability density & 0.070 \\ 
\hline
\end{tabular}
\end{center}
\label{tab:security_results}
\end{table}%

Obviously, the prediction is not very accurate, but the distribution shifts in the
correct direction as the mean of the distribution used for the vulnerability density
is 0.0054. This effect is similar as in the CM1 case in section~\ref{sec:maintainability_cases}.
The model cannot completely overcome deficiencies in the underlying distributions
for the indicators.

Windows and Linux have 15--40 MSLOC, i.e., the are an order of
magnitude larger than Tomcat. Hence, one explanation of the large difference
could be that the used underlying distribution for
the vulnerability is wrong and need to be improved with data from systems with
other sizes. Another possibility is that the system size is a fact that needs to be
considered. In this case, it seems that smaller systems have a higher vulnerability
density. This could be explained by the fact that in Tomcat almost all parts of the
system are exposed to external access. Therefore, the fact \emph{Size of Software}
might be introduced as an additional factor. However, experimenting with this additional
fact showed that it cannot reach accurate prediction as it is still constrained
by the vulnerability density distribution.

The large difference can also mean that our chosen metric \emph{vulnerability density} is not
suitable as it seems to be strongly dependent on the system size, type, or both. Hence,
more advanced metrics such as \emph{breach rates} or \emph{cost to breach}
\cite{schlechter05} should be used. These measures are, however, significantly
harder to measure.

\subsection{Discussion}

The foremost goal of these examples was not to show predictive
validity but to investigate the general feasibility of the assessment and
prediction approach. Nevertheless, in two of the maintainability cases, we were
able to come to accurate predictions although we had little knowledge about the
actual system. Furthermore, we saw that it is possible to build a Bayesian model with
reasonable effort. The four-step approach gives direct
guidance for most of the network building. However, setting up the NPTs is still a
challenge. There are usually several possibilities how a relationship can be
expressed and with how much uncertainty it should be afflicted. This needs
expert opinion and experimentation. Nevertheless, AgenaRisk provides
sophisticated tool support to find easier ways to define an NPT. 

Another problem encountered in the analysed cases was to find a reasonable empirical
data basis for the distribution of the indicators. For the facts and activities, it is
sufficient to employ rather coarse-grained ranked states and the good tool support
helps in defining corresponding distributions. For the indicators, a good data basis
or distributions from the literature are crucial.

As our ABQMs can get
very large with a few hundreds of model elements, it remains to be evaluated
whether the approach scales when the quality model is fully mapped to a
Bayesian network. Probably a selection of a subset of the quality model is
necessary first.

Furthermore, it is important to note that a more in-depth validation of a
resulting Bayesian network is necessary in order to ensure that all parts --
topology, node states, and NPTs -- represent the interdependencies of
the quality factors good enough so that a valid statement about the quality
of the software system can be made. This is not covered by these examples 
but has to be the next step.

The usage of the proposed approach, however, has benefits beyond assessing
a quality goal. If we are able to calibrate the model and establish sufficient
accuracy in the Bayesian network, we will be able to calculate low level goals for
quality goals. For example, we can infer requirements for the indicators of the
maintainability cases for a given goal for the average change effort. A required
average change effort of only 10 person-hours has the most likely explanation in
a comment ratio of 0.3, an average cyclomatic complexity of 6.4, and an average
module size of 64 LOC. This can be used to guide development. Finally, also
requirements specification is supported by this approach as quality requirements
now can use the given indicators to specify assessable requirements.

\section{Related Work}
\label{sec:related}

The basic idea to use Bayesian networks for assessing and predicting software
quality has been developed foremost by Fenton, Neil, and Littlewood. They
introduced Bayesian networks as a tool in this area and applied it in various
contexts related to software quality. In \cite{fenton99} they formulate a critique
on current defect prediction models and suggest to use Bayesian networks.
Other researchers also used Bayesian networks for software quality prediction
similarly \cite{amasaki05,perezminana06}

The work closest to the approach proposed in this paper is \cite{neil99}.
The authors discuss quality models such as the ISO 9126 \cite{iso9126-1:2001}
and their problems such as the undefinedness of the relationships in such
a model. They aim at solving these problems by defining Bayesian networks
for quality attributes directly. Our work differentiates in using a defined structure
for quality models that contain far more details as common quality models.
This structure and detail allows a straightforward derivation of a Bayesian
network from the quality model. This has the advantage that the basic quality
model can also be used for other purposes then prediction such as the
specification of quality requirements.

Beaver, Schiavone, and Berrios \cite{beaver05} also used a Bayesian network
to predict software quality including diverse factors such as team skill and
process maturity. In his thesis \cite{beaver06}, Beaver even compared the
approach to neural networks and Least Squares regressions that both were
outperformed by the Bayesian network. However, they did not rely on a
structured quality model as in our approach.

An earlier version of this paper was published in \cite{wagner:promise09}. It
already contained the 4-step approach as described in this paper. However,
we added 3 additional projects to the maintainability case and a completely
new security case in order to validate the applicability of the approach more
broadly.

\section{Conclusions}\label{sec:conclusions}

A high goal in software quality management is the reliable quantitative
assessment and prediction of software quality. Many efforts
in building assessment and prediction models have given insights
in the usefulness but also the constraints of such models. However, these
models have not been tightly integrated into other quality management
activities. Activity-based quality models have proven in practice to be a
solid foundation for defining quality on a detailed level. However, quantitative
analyses have not been directly possible so far.

Bayesian networks have been shown to provide promising results in
quality predictions. Because of that and their clear structuring that can
straightforwardly reflect the structure of activity-based quality models, a
four-step approach for transferring activity-based quality models to Bayesian
networks was proposed. It allows to systematically construct a Bayesian
network that uses the knowledge encoded in the quality model to provide
information about a given assessment or prediction goal. In the terminology of
\cite{deissenb:wosq09}, we use a \emph{quality definition model} (the ABQM)
and enrich it with a \emph{quality assessment} and \emph{quality prediction
model} (the Bayesian Network).

Although not fully validated, we demonstrated the approach in a
study using real NASA project data and automatically measured bug patterns
of the Tomcat servlet container. The examples showed the
applicability of the approach to such projects and could improve
the prediction in comparison to industry standard values. For 2 of the 5 cases,
the predictions were accurate despite the usage of literature values as basis
for the distributions only. 

The use of Bayesian networks opens many possibilities. Most interestingly,
after building a large Bayesian network, a sensitivity analysis of that
network can be performed. This can answer the practically very relevant
question which of the factors are the most important ones. It would
allow to reduce the measurement efforts significantly by concentrating
on these most influential facts.

We plan to apply this approach in future case studies at our industrial
partners in order to further validate the approach. A comprehensive
analysis of the predictive validity is necessary to judge the usefulness
of the approach and to compare it with other means for assessment
and prediction.

\section{Acknowledgments}

This work has partially been supported by the German Federal Ministry 
of Education and Research (BMBF) in the project QuaMoCo (01 IS 08023B). 
I would like to thank the anonymous reviewers
at PROMISE for feedback on an earlier version and Vic Basili,
Sebastian Winter, and the anonymous IST reviewers
 for helpful suggestions for this version.

\balance
%\bibliographystyle{elsarticle-num}
%\bibliography{ist_qa}

\end{document}